# Oxygen vacancy driven mobility enhancement in epitaxial La-doped BaSnO$_3$ from vacuum annealing

Hai Jun Cho[1,2, a)], Takaki Onozato[2], Mian Wei[2], Anup Sanchela[1],

and Hiromichi Ohta[1,2, a)]

[1]Research Institute for Electronic Science, Hokkaido University, N20W10, Kita, Sapporo 001−0020, Japan

[2]Graduate School of Information Science and Technology, Hokkaido University, N14W9, Kita, Sapporo 060−0814, Japan

**Corresponding authors**

[a)]Correspondence and requests regarding this manuscript should be addressed to:

Email: joon@es.hokudai.ac.jp, hiromichi.ohta@es.hokudai.ac.jp

**Wide bandgap (~3.1 eV) La-doped BaSnO$_3$ (LBSO) has attracted increasing attention as one of the transparent oxide semiconductors since its bulk single crystal shows a high carrier mobility (~320 cm$^2$ V$^{-1}$ s$^{-1}$) with a high carrier concentration (~10$^{20}$ cm$^{-3}$). For this reason, many researchers have fabricated LBSO epitaxial films thus far, but the obtainable carrier mobility is substantially low compared to that of single crystals due to the formation of the lattice/structural defects. Here we report that the mobility suppression in LBSO films can be lifted by a simple vacuum annealing process. The vacuum annealing of the LBSO films on MgO substrate increased the carrier concentrations due to the oxygen vacancy formation, which leads to simultaneous lateral grain growth. As a result, the carrier mobility was greatly improved by the vacuum annealing, which does not occur after heat treatment in air. These results expand our current knowledge on the point defect formation in epitaxial LBSO films and show that vacuum annealing is a powerful tool for enhancing the mobility values of LBSO films.**

Transparent oxide semiconductors (TOSs) are promising candidates for various future electronic devices such as transistors, solar cells, and display panels.[1,2] For such applications, high electron mobility ($\mu$) is essential, and this has been a great disadvantage for TOS materials since their mobility values are low compared to classical semiconductors. In this regard, perovskite La-doped BaSnO$_3$ (LBSO) is gaining significant interest since its bulk single crystal exhibits a wide bandgap (~3.1 eV) and a very high mobility of 320 cm$^2$ V$^{-1}$ s$^{-1}$,[3,4] which is comparable to that of single crystal Si (~350 cm$^2$ V$^{-1}$ s$^{-1}$).[5] For this reason, there have been many attempts to utilize LBSO in thin films transistors.

However, crystalline defects prevent the $\mu$ in LBSO films from reaching the single crystal value, and many studies were devoted to improving the crystal quality of epitaxial LBSO films.[6-12] For example, since misfit dislocations occur at the film/substrate interface due to the lattice mismatch, buffer layers are commonly used to reduce the dislocations.[13, 14] To completely eliminate the film/substrate mismatch, Lee *et al.* grew LBSO films on single crystal BaSnO$_3$ substrate.[15] In addition, in our recent study, we fabricated LBSO films under an ozone atmosphere to reduce the amount of point defects.[16,17] Unfortunately, while these approaches were successful in improving the mobility values in LBSO films (up to ~120 cm$^2$ V$^{-1}$ s$^{-1}$), they can significantly increase the fabrication cost, which may be a crucial issue in mass production systems at industrial scales.

In large scale production facilities, designing a clever post treating process is often more economical than improving the quality of as-deposited samples. In this regard, very interesting experimental results were released in 2015 and 2018. In one study, N$_2$ environment at 1000 °C was used to create oxygen vacancies in the LBSO films on SrTiO$_3$ substrates, and the $\mu$ increased from 41 cm$^2$ V$^{-1}$ s$^{-1}$ to 78 cm$^2$ V$^{-1}$ s$^{-1}$.[18] In the other study (same research group), oxygen vacancies were generated in the LBSO films on SrTiO$_3$ substrates using H$_2$ forming gas at 950 °C, which further improved the $\mu$ up to 122 cm$^2$ V$^{-1}$ s$^{-1}$.[19] According to these studies, oxygen vacancies can neutralize the negative charges at threading dislocations.[20-22]

As removing oxygen ions (O$^{2-}$) near threading dislocations decreases their thermal stability, oxygen vacancy doping creates a very strong driving force for lateral grain growths at high temperatures, which significantly increases the free propagation length of the carrier electrons. These results suggest that post treating of LBSO films can be just as effective as modifying the synthesis methods to improve the crystal quality of the as deposited LBSO films. In undoped BaSnO$_3$ films, vacuum annealing is commonly used to create oxygen vacancies and induce mobile charge carriers.[23, 24] Since vacuum annealing process is much simpler than creating N$_2$ or H$_2$ forming gas environment, it can be an alternative method for inducing oxygen vacancy assisted grain growths in LBSO films. There is one study that examined the effect of vacuum annealing on the electron transport properties of LBSO films,[25] but the reported $\mu$ values are too low (< 4 cm$^2$ V$^{-1}$ s$^{-1}$) to validate vacuum annealing as an effective method for improving $\mu$.

For optimizing the post annealing process, it is important to understand point defect formation in LBSO films because the $\mu$-enhancement in LBSO films is triggered by oxygen vacancies. In stoichiometric BaSnO$_3$, the oxidation states of its constituents are: Ba$^{2+}$ = [Xe], Sn$^{4+}$ = [Kr] 4d$^{10}$, and O$^{2-}$ = [Ne]. Since two of the constituents (Ba$^{2+}$, O$^{2-}$) exhibit the same orbitals with inert gases (Xe, Ne), stoichiometric BaSnO$_3$ crystals are believed to be thermodynamically stable,[26] and stoichiometric BaSnO$_3$ films do not intrinsically conduct electricity as all electrons form firmly bound states. The formation energy of oxygen vacancy in BaSnO$_3$ is high and can only be lowered by reducing the chemical potential of oxygen,[27] which can be achieved by lowering oxygen pressure during the film growth[28, 29] or vacuum annealing at high temperatures.[23, 24] Therefore, as-deposited BaSnO$_3$ films do not have sufficient oxygen vacancies to conduct electricity unless they are intentionally created. In contrast, oxygen vacancies are much more common in LBSO films even if sufficient oxygen is provided during the film growth.[19] In one of our previous studies, Sn$^{2+}$ states were detected from LBSO films fabricated under 10 Pa of O$_2$, which implies the presence of oxygen deficiency.[16, 28] These results suggest that the La-dopants may be promoting oxygen vacancy formations in BaSnO$_3$ films. However, the relationship between La-dopants and oxygen vacancy in BaSnO$_3$ has not been investigated in detail.

In this study, we studied the effect of La-dopants on the formation of oxygen vacancy in LBSO films and investigated the feasibility of enhancing the $\mu$ of LBSO films using vacuum annealing. We expected that the vacuum annealing can trigger oxygen vacancy induced grain growth in LBSO films, and the resulting mobility enhancement can depend on the doping level of La ([La$^{3+}$]). According to our results, vacuum annealing significantly increased the $\mu$ of LBSO films. We also found that La-dopants increased the oxygen vacancy vs. lattice oxygen ($V_O/L_O$) ratio in as-deposited LBSO films and affected the vacuum annealing effect on LBSO films. The results of this study expand our current knowledge on the point defect formation in epitaxial LBSO films and show that vacuum annealing is a simple and effective method for enhancing the electron mobility of LBSO films.

LBSO epitaxial films ($[La^{3+}] = 0.1, 0.55, 1, 2, 5$, and 7 %) were fabricated on (001)-oriented MgO substrates at 750 °C using pulsed laser deposition (PLD, KrF excimer laser, fluence ~2 J cm$^{-2}$ pulse$^{-1}$, repetition rate = 10 Hz). The oxygen pressure inside the chamber during the film growth was kept at 10 Pa. High-resolution X-ray diffraction (XRD, Cu K$\alpha_1$, ATX-G, Rigaku Co.) and reciprocal space mappings (RSMs) were performed around (204) diffraction spots of the LBSO films. The film thicknesses were measured from the Kiessing or Pendellöesung fringes in the XRD patterns (data not shown). The electrical conductivity ($\sigma$), carrier concentration ($n$), and Hall mobility ($\mu_{Hall}$) of the films were measured at room temperature by the conventional dc 4-probe method in van der Pauw electrode geometry.

To confirm that vacuum annealing can induce grain growth in LBSO films, a 2 % LBSO film (~64 nm) was prepared and cut it into two pieces. We annealed one piece in air and the other piece in vacuum ($< 10^{-2}$ Pa) at 750 °C for 1 h. Then, we measured the lateral grain sizes of the films before and after the heat treatments using the RSM [**Fig. 1**]. While the (204) diffraction spot of the air annealed film was almost the same with that of as-deposited film, the (204) diffraction spot of the vacuum annealed was two times more intense compared to the other two films (as-deposited & air-annealed). The lateral grain sizes ($D$) of the as-deposited, air annealed, and vacuum annealed LBSO films were 9.1 nm, 10.2 nm, and 22.3 nm, respectively. The small grain size change after air annealing is not surprising since the film was deposited at 750 °C, and the no significant change in the microstructure was expected. On the other hand, the vacuum annealing substantially increased the lateral grain size, which is consistent with the H$_2$ forming gas experiment.[19] This shows that vacuum annealing can indeed be an alternative method for triggering oxygen vacancy assisted grain growth in LBSO films.

In order to find the optimum vacuum annealing temperature, several 2 % LBSO films with similar thicknesses (~41 nm) and electrical properties were fabricated (**Supplementary Table S1**), and they were annealed in vacuum ($< 10^{-2}$ Pa) at different temperatures ranging from 650 °C to 800 °C for 30 min. With increasing annealing temperature, the $D$ of the films increased gradually from ~5 nm to ~17 nm [**Fig. 2(a)** and **Supplementary Fig. S1**]. **Figure 2(b) and 2(c)** summarize the effect of the vacuum annealing on the electron transport properties of the LBSO film. The $n$ increased up to ~$1.8 \times 10^{20}$ cm$^{-3}$ from ~$1.35 \times 10^{20}$ cm$^{-3}$ when the sample was annealed in the vacuum [**Fig. 2(b)**]. In addition, the observed $n$ was far lower than the [La$^{3+}$] (=$2.87 \times 10^{20}$ cm$^{-3}$), indicating the low carrier activation rate of the La$^{3+}$ ions. On the other hand, the $\mu_{Hall}$ greatly increased (up to ~80 cm$^2$ V$^{-1}$ s$^{-1}$) with the vacuum annealing temperature. The highest $\mu_{Hall}$ was observed from the LBSO film vacuum annealed at 725 °C [**Fig. 2(c)**]. Above 725 °C, the $\mu_{Hall}$ starts to decrease despite the increase in the $D$, suggesting the oxygen vacancy generation rate is too high. Therefore, we decided 725 °C to be the optimal vacuum annealing temperature.

Interestingly, the vacuum annealing effect showed a strong thickness dependence as the mobility enhancement was noticeably larger in thinner films (**Supplementary Figs. S2 and S3**). Since the lateral grain growth in LBSO films can only take place if oxygen vacancies are supplied, the annealed region is limited by the diffusion of oxygen vacancy from the outer surface. Therefore, the observed film thickness dependence implies that mobility suppression in the LBSO films starts from the film/substrate interface, which is consistent with one of our earlier studies on the thickness dependence of the carrier mobility in LBSO.[17] The highest mobility observed was 101.6 cm$^2$ V$^{-1}$ s$^{-1}$ from vacuum annealed 117 nm LBSO film, which is comparable to that observed in LBSO films with buffer layers.

To find the effect of La-dopants on the vacuum annealing process, we annealed LBSO films (average thickness: 43 nm) with varying [La$^{3+}$] at 725 °C in vacuum for 30 min. **Figure 3** summarizes the electron transport properties of the annealed LBSO films. The

$n$ of the LBSO films always showed an increase with [La$^{3+}$] while the vacuum annealed films exhibited higher $n$ compared to the as-deposited films [**Fig. 3(a)**]. The 0.1 % LBSO film, which did not conduct electricity, also became electrically conductive. The absolute value of thermopower ($S$) decreased with [La$^{3+}$], consistent with the relationship between $n$ and [La$^{3+}$] [**Fig. 3(b)**].[30] The $\mu_{Hall}$ of the annealed LBSO films was dramatically enhanced compared with the as-deposited LBSO films [**Fig. 3(c)**]. Interestingly, the $\mu_{Hall}$ enhancement was much greater in films with ≤ 2 % doping compared to that in films with ≥ 5 % doping. While the highest as-deposited $\mu_{Hall}$ was observed in the 5 % LBSO film, the highest $\mu_{Hall}$ after vacuum annealing was observed in the 2 % LBSO film (35.5 cm$^2$ V$^{-1}$ s$^{-1}$ → 72.3 cm$^2$ V$^{-1}$ s$^{-1}$). Furthermore, while the annealed lateral grain size increased with [La$^{3+}$], and the lateral grain size enhancement did not change after 2% [La$^{3+}$] [**Fig. 3(d)** and **Supplementary Fig. S4**]. These results indicate that 2 % LBSO films have an optimized balance between oxygen vacancy formation ($\mu_{Hall}$ ↓) and oxygen vacancy induced grain growth ($\mu_{Hall}$ ↑) during vacuum heat treatments.

In order to further investigate the effect of the vacuum annealing on the $\mu$ enhancement of the LBSO films, we performed the X-ray photoelectron spectroscopy (XPS) measurement of the LBSO films [**Fig. 4(a): as-deposited, Fig. 4(b): Vacuum annealed**]. In case of oxygen in oxides, lattice oxygen peak (L$_O$) is located at ~529 eV. If oxygen vacancies are present, this peak shift to ~531 eV ($V_O$).[19,31,32] Another oxygen peak around ~532 eV (A$_O$) can emerge from chemically adsorbed oxygen from surface contamination by organic molecules. The source of chemically adsorbed oxygen is unknown, but we believe it is related to the status of the vacuum chambers (annealing, XPS). For the XPS peak fitting, a convolution between Gaussian (70 %) and Lorentzian (30 %) was used. For each LBSO films, the full width at half maximum (FWHM) was constrained to be the same for all 3 oxygen peaks.

The $V_O$ peak energies of the LBSO films increased after vacuum annealing, especially for films with higher [La$^{3+}$] [**Fig. 4(c)**]. The $V_O$/L$_O$ area ratio, which was extracted from the XPS data (**Supplementary Table S2**), in the as-deposited LBSO films increased gradually when the [$La^{3+}$] exceeded 1 %. Upon vacuum annealing, this ratio increased further for all films except for 7 % LBSO film. The largest change in $V_O/L_O$ was observed from 2% doped LBSO film. Since the solubility limit of La in LBSO was reported to be ~5 %, the oxygen vacancy reduction in 7 % LBSO is likely attributed to the formation of $La_2Sn_2O_7$,[33] which can be observed from the ceramic targets used to deposit the films (**Supplementary Fig. S5**). These results show that La-dopants in epitaxial LBSO films affect not only the oxygen stability, but also the overall thermal stability of the film.

Unfortunately, although we successfully increased the $\mu$ of the LBSO films, the role of La-dopants in the oxygen vacancy formation mechanism in LBSO films is still unclear. In this regard, we believe that the role of threading dislocation in point defect formation is important. For example, in case of unintentionally $V_O$ doped $BaSnO_{3-\delta}$ single crystal, vacuum annealing reduces the carrier concentration and therefore reduces the oxygen vacancy level.[34] This contradicts the behavior of epitaxial $BaSnO_3$ films, where vacuum annealing increases the carrier concentration.[22, 23] Since the main structural difference between single crystals and epitaxial films is the presence of threading dislocations, it is plausible to think that they can promote point defect formation in epitaxial films. In the context of this research, since impurities often segregate between grains separated by dislocations,[35] one possibility is the segregation of La-dopants at threading dislocations, which is plausible since $La^{3+}$ ions can compensate the missing cation charges at threading dislocations.[20-22] This scenario also explains the low dopant carrier activation rate observed in epitaxial LBSO films [**Fig. 2(a)**]. In this case, $La^{3+}$ vacant sites in the grain interior may lose adjacent $O^{2-}$ ions due to the lack of bonding electrons.

In summary, we examined the effect of vacuum annealing on the electron transport properties of the epitaxial LBSO films on (001) MgO substrates. The results clearly showed that the vacuum annealing is powerful technique to enhance the carrier mobility of LBSO films. We also found the oxygen vacancy vs. lattice oxygen ($V_O/L_O$) ratio in the vacuum annealed LBSO films was almost double as compared with the as-deposited LBSO films. In addition, lateral grain size of the LBSO films substantially increased after vacuum annealing whereas it remained almost unchanged after air annealing. These results clearly show that vacuum annealing triggers oxygen vacancy induced grain growth in the LBSO films.

The vacuum annealing approach was very effective for films with small thicknesses. Therefore, it is a very good method for making LBSO film transistors since low thicknesses are desired for reducing the power consumption. We believe these results will be useful for designing low cost fabrication methods for high-mobility LBSO films or can be used to improve the carrier mobility of other perovskite stannates such as $SrSnO_3$.

## Acknowledgments

This research was supported by Grants-in-Aid for Scientific Research A (17H01314) from the JSPS, the Asahi Glass Foundation, and the Mitsubishi Foundation. A part of this work was supported by Dynamic Alliance for Open Innovation Bridging Human, Environment and Materials as well as the Network Joint Research Center for Materials and Devices.

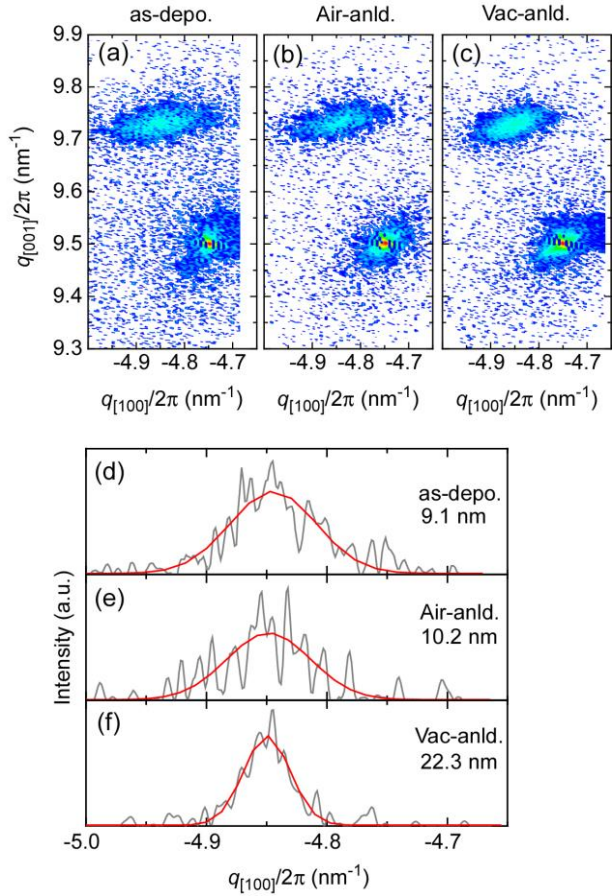

**FIG. 1 | Lateral grain growth of the LBSO film by the vacuum annealing.** (a)−(c) RSMs near (204) diffraction peak of 2 % La-doped LBSO films (~64 nm) [(a) as-deposited film, (b) air-annealed at 750 °C for 1 h, and (c) vacuum annealed at 750 °C for 1 h]. The RSMs were shifted using the peak position of (204) MgO ($q_{[100]}/2\pi = 9.50$ nm$^{-1}$, $q_{[100]}/2\pi = -4.75$ nm$^{-1}$). (d)−(f) Cross-sectional profiles of the LBSO peaks (a)−(c). The $D$ values extracted from the RSMs were (a)(d) 9.1 nm, (b)(e) 10.2 nm, and (c)(f) 22.3 nm, respectively. The as-deposited film was cut into two pieces. One was annealed in air while the other was annealed in vacuum. Almost no change in the RSM was observed from air-annealing. On the other hand, significant increase in the lateral grain size and peak intensity can be noticed from the vacuum annealed film.

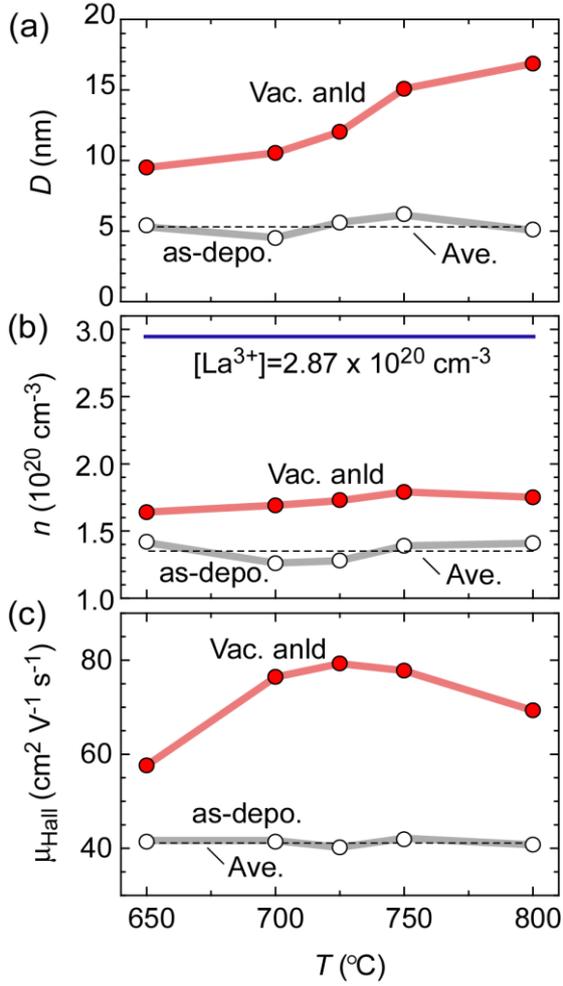

**FIG. 2 | Optimum vacuum annealing temperature for the 2 % La-doped LBSO film.** (a) Lateral grain size ($D$) increased gradually with vacuum ($<10^{-2}$ Pa) annealing temperature. (b) The carrier concentration ($n$) increased up to ~$1.8\times10^{20}$ cm$^{-3}$ from ~$1.35\times10^{20}$ cm$^{-3}$ when the sample was annealed in the vacuum. The observed $n$ is lower than [La$^{3+}$] (=$2.87\times10^{20}$ cm$^{-3}$), indicating the activation of La$^{3+}$ is low. (c) The Hall mobility ($\mu_{Hall}$) greatly increased with temperature when the sample was annealed in the vacuum up to ~80 cm$^2$ V$^{-1}$ s$^{-1}$ from the as-deposited mobility of ~40 cm$^2$ V$^{-1}$ s$^{-1}$. From these results, we decided the optimum vacuum annealing temperature (725 °C).

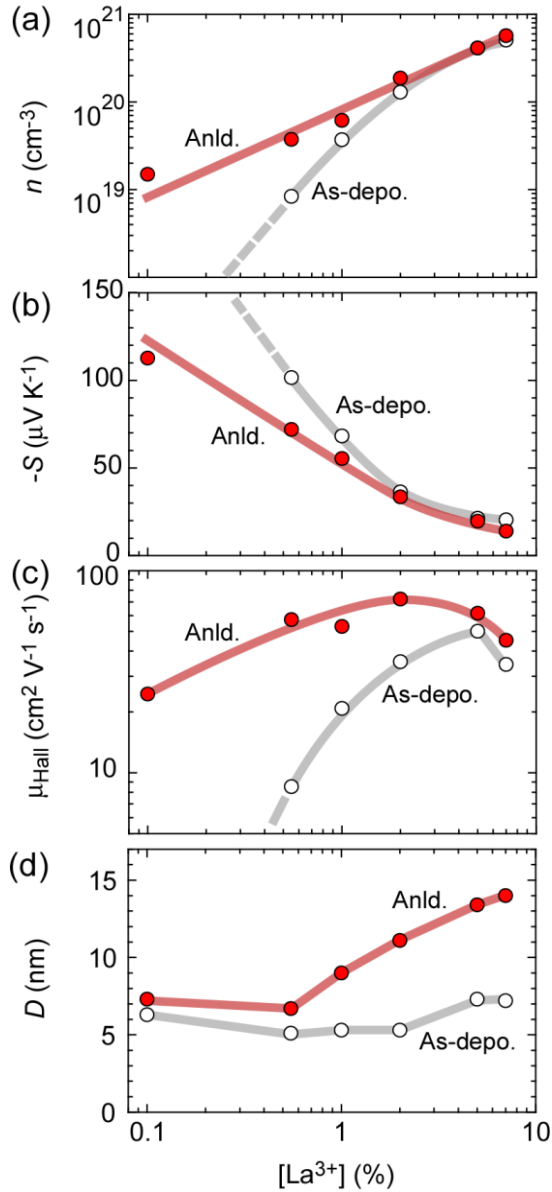

**FIG. 3 | La-concentration dependences of the electron transport properties and the lateral grain size for the vacuum annealed LBSO films (725 °C, <10$^{-2}$ Pa, 30 min).** (a) Carrier concentration ($n$) of the vacuum annealed LBSO films shows linearly increases with [La$^{3+}$] while the as-deposited LBSO films show lower $n$. (b) The absolute value of thermopower ($S$) decreases with [La$^{3+}$], consistent with the $n$ vs [La$^{3+}$] relation.[30] (c) The Hall mobility ($\mu_{Hall}$) of the annealed LBSO films was dramatically enhanced at low [La$^{3+}$] compared with the as-deposited LBSO films. (d) Lateral grain size ($D$) of the LBSO films before and after vacuum annealing. $D$ for the annealed samples increased when [La$^{3+}$] is greater than 1 %.

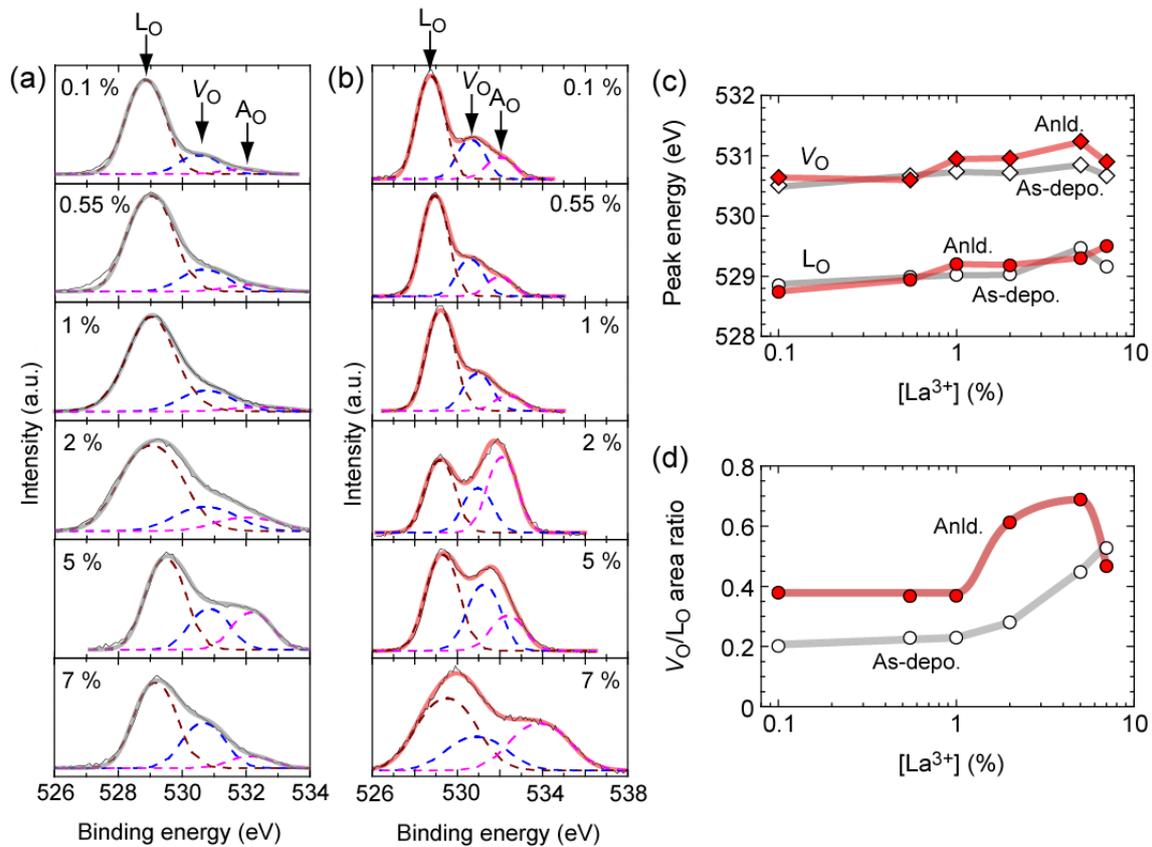

**FIG. 4 | Oxygen vacancy evolution in the LBSO films after the vacuum annealing.** XPS spectra of the LBSO films at (a) the as-deposited states and (b) after the vacuum annealed states. The XPS peaks were decomposed into three peaks [$L_O$: Lattice oxygen (~529 eV), $V_O$: Oxygen vacancy (~531 eV), and $A_O$: chemically adsorbed oxygen (~532 eV)]. (c) Peak energies of $L_O$ and $V_O$ at different [$La^{3+}$]. The $V_O$ at higher [$La^{3+}$] of the annealed sample locates higher energies as compared with that of as-deposited samples. (d) $V_O/L_O$ area ratio, which was extracted from the XPS data. The $V_O/L_O$ ratio in the as-deposited LBSO films increases gradually when the [$La^{3+}$] exceed 1 %. The $V_O/L_O$ ratio in the annealed LBSO films is almost double as compared with the as-deposited LBSO films.

**Supplementary Information**

# Oxygen vacancy driven mobility enhancement in epitaxial La-doped BaSnO$_3$ from vacuum heat treatments

Hai Jun Cho[1,2, a)], Takaki Onozato[2], Mian Wei[2], Anup Sanchela[1],

and Hiromichi Ohta[1,2, a)]

[1]Research Institute for Electronic Science, Hokkaido University, N20W10, Kita, Sapporo 001− 0020, Japan

[2]Graduate School of Information Science and Technology, Hokkaido University, N14W9, Kita, Sapporo 060−0814, Japan

**Corresponding authors**

a)Correspondence and requests regarding this manuscript should be addressed to:

Email: joon@es.hokudai.ac.jp, hiromichi.ohta@es.hokudai.ac.jp

**Table S1 | Properties of as-deposited 2 % La-doped LBSO films for the vacuum annealing experiments.** $T$: Temperature for the vacuum annealing, $t$: Film thickness, $\sigma$: Electrical conductivity, $\mu_{Hall}$: Hall mobility, $D$: Lateral grain size

| $T$ (°C) | $t$ (nm) | $\sigma$ ($10^1$ S cm$^{-1}$) | $n$ ($10^{20}$ cm$^{-3}$) | $\mu_{Hall}$ (cm$^2$ V$^{-1}$ s$^{-1}$) | $D$ (nm) |
|---|---|---|---|---|---|
| 650 | 42.1 | 93.6 | 1.42 | 41.4 | 5.4 |
| 700 | 39.3 | 83.5 | 1.26 | 41.4 | 4.5 |
| 725 | 38.6 | 82.6 | 1.28 | 40.2 | 5.6 |
| 750 | 43.2 | 93.6 | 1.29 | 41.9 | 6.2 |
| 800 | 42.0 | 92.3 | 1.41 | 40.8 | 5.1 |
| Average | 41.0 ±2.0 | 89.1 ±5.5 | 1.35 ±0.07 | 41.1 ±0.7 | 5.3±0.9 |

**Table S2 | O 1s XPS peak areas before and after vacuum annealing.**

| [La$^{3+}$] (%) | As deposited | | Vacuum annealed | |
|---|---|---|---|---|
| | L$_O$ | $V_O$ | L$_O$ | $V_O$ |
| 0.1 | 2692 | 543 | 1683 | 637 |
| 0.55 | 2517 | 576 | 1297 | 477 |
| 1 | 2654 | 608 | 1927 | 714 |
| 2 | 1544 | 432 | 740 | 453 |
| 5 | 1176 | 526 | 1874 | 1289 |
| 7 | 1756 | 925 | 1343 | 627 |

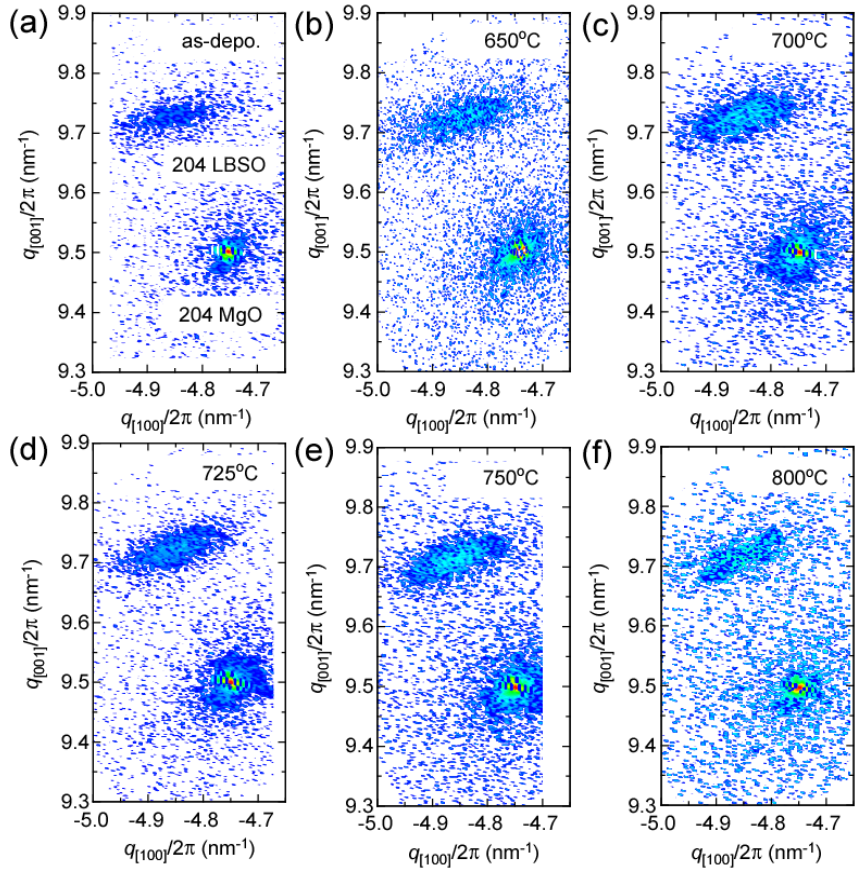

**FIG. S1 | RSMs near (204) peak of 2% LBSO (~41 nm) films before and after vacuum annealing at different temperatures.** (a) as-deposited, (b) 650 °C, (c) 700 °C, (d) 725 °C, (e) 750 °C, and (f) 800 °C. The RSMs were shifted using the peak position of (204) MgO ($q_{[100]}/2\pi = 9.50$ nm$^{-1}$, $q_{[100]}/2\pi = -4.75$ nm$^{-1}$). The $D$ values extracted from the RSMs were (a) 5.3 nm, (b) 9.5 nm, (c) 10.5 nm, (d) 12.0 nm, (e) 15.1 nm, and (f) 16.9 nm, respectively.

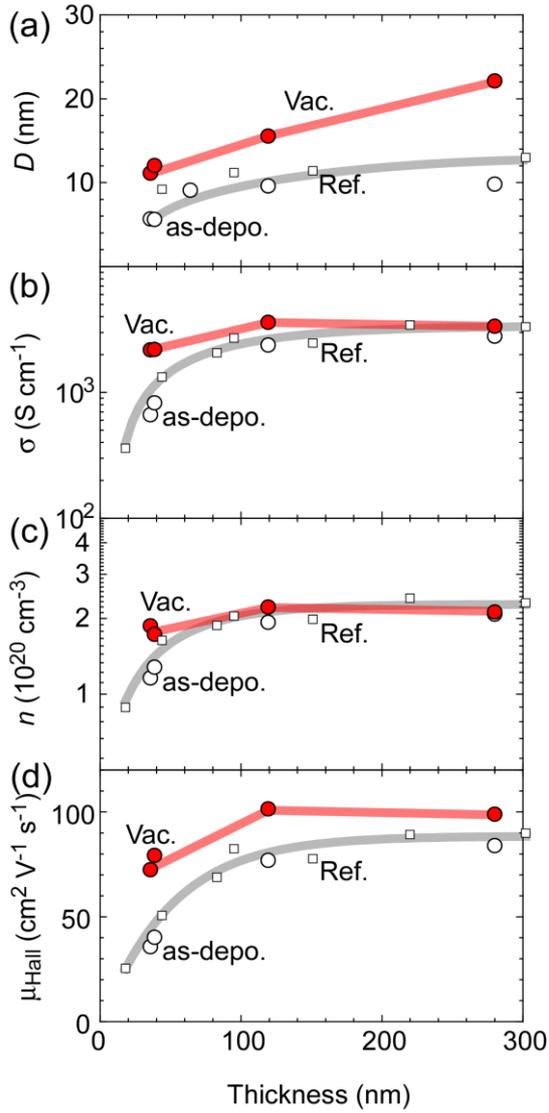

**FIG. S2 | Thickness dependences of the electron transport properties for the vacuum annealed LBSO films at 725 °C.** (a) Lateral grain size ($D$), (b) electrical conductivity ($\sigma$), (c) carrier concentration ($n$), and (d) Hall mobility ($\mu_{Hall}$) of 2 % LBSO films [Red circles: vacuum annealed, white circles: as-deposited, white squares: as-deposited properties from our previous study (Ref. 17)[17]]. The vacuum annealing effects are noticeably larger in thin films. This suggests that the mobility suppression in epitaxial LBSO films originates from the interface, and the oxygen vacancy assisted grain growth is limited by the diffusion of oxygen vacancies.

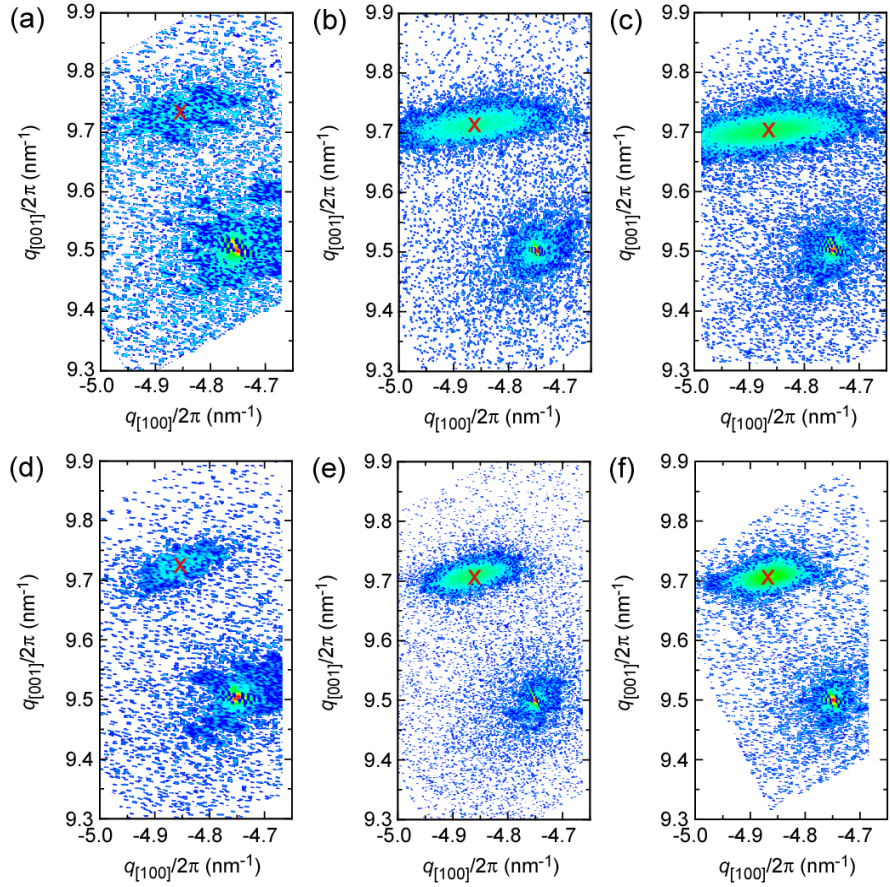

**FIG. S3 | RSM near (204) peaks of 2 % LBSO films with different thickness before (a−c) and after vacuum annealing (d−f) at 725 °C.** The thickness were (a)(d) 36 nm, (b)(e) 117 nm, and (c)(f) 280 nm. The RSMs were shifted using the peak position of (204) MgO ($q_{[100]}/2\pi = 9.50$ nm$^{-1}$, $q_{[100]}/2\pi = -4.75$ nm$^{-1}$). The $D$ values extracted from the RSMs were (a) 5.7 nm, (b) 9.6 nm, (c) 9.8 nm, (d) 11.1 nm, (e) 15.6 nm, and (f) 22.1 nm, respectively.

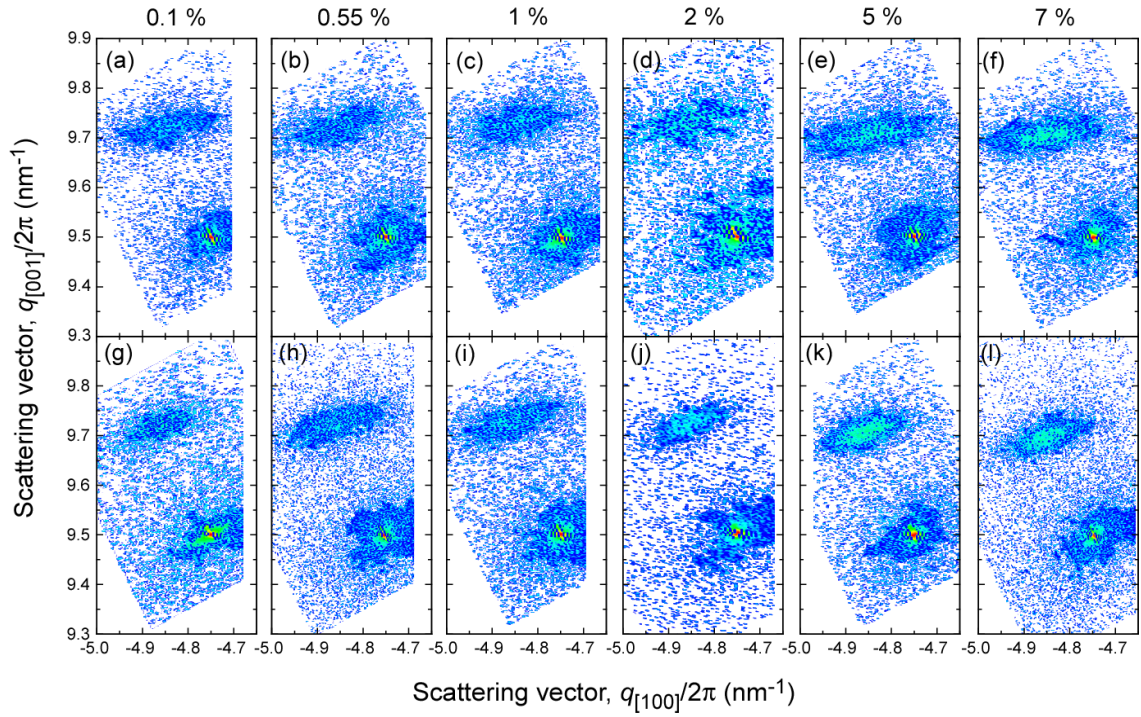

**FIG. S4 | RSM near (204) peaks of LBSO (~43 nm) films with different [La$^{3+}$] before (a–f) and after vacuum annealing (g–l) at 725 °C.** The RSMs were shifted using the peak position of (204) MgO ($q_{[100]}/2\pi = 9.50$ nm$^{-1}$, $q_{[100]}/2\pi = -4.75$ nm$^{-1}$). The $D$ values extracted from the RSMs were (a) 6.3 nm, (b) 5.1 nm, (c) 5.3 nm, (d) 5.3 nm, (e) 7.3 nm, (f) 7.2 nm, (g) 7.3 nm, (h) 6.7 nm, (i) 9.0 nm, (j) 11.1 nm, (k) 13.4 nm, and (l) 14.0 nm, respectively.

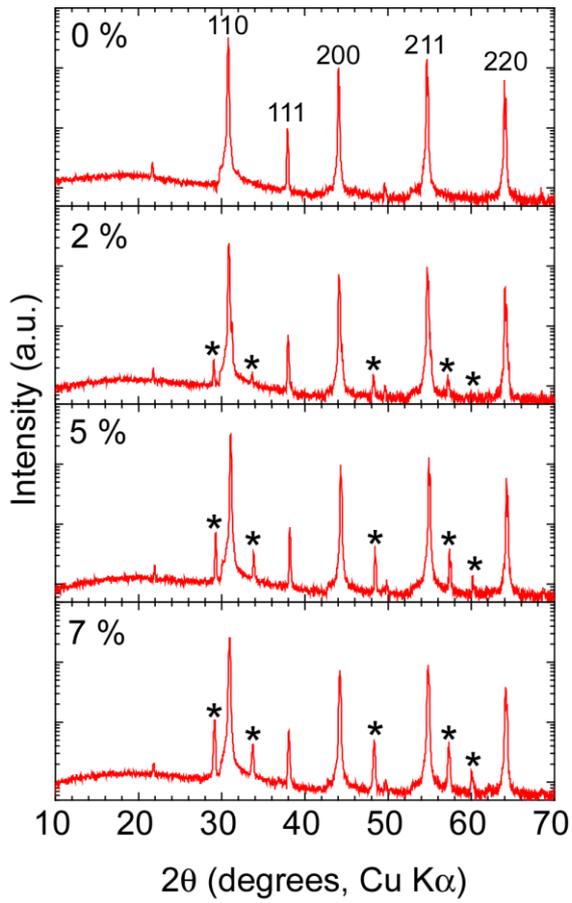

**FIG. S5 | Powder XRD patterns of the LBSO ceramic targets used for the LBSO film growth.** As the [$La^{3+}$] increases, peaks from $La_2Sn_2O_7$ (*) emerges.